# Proposal for a Photonic Remote Active Heat Sink Technology (PHRAHST)

Dimitris Dimitropoulos, Bahram Jalali

Optoelectronic Circuit & Systems Laboratoty

University of California, Los Angeles, CA90095

**Abstract –** We propose a new method to effect heat removal from an object by using a laser beam. The proposed method is based on inelastic scattering of a laser beam from the object and in particular by making the anti-Stokes emission more efficient than the Stokes emission. In that manner more energy is removed from the body per unit time than deposited. Various ways are outlined in order to achieve this result ranging from careful selection of the laser frequency with respect to the resonant frequencies of the medium, use of the frequency dependence of the density of electromagnetic modes in a three-dimensional system, use of photonic crystals and the polarization dependence of electromagnetic modes in cavities. The proposed methods could find use for example in the cooling of devices of nanoscale dimensions.

1. **Introduction**

Heat removal from optical and electronic devices is of central importance since heating limits the performance and utility of these devices. The problem of heating is most pronounced in current CMOS microprocessors where it has been identified as the most important challenge facing the industry [1]. The continuing trend of scaling down device dimensions to the nanometer regime increases the utility of present invention as cooling down nanoscale devices by conventional methods used for macroscopic dimensions will be inefficient. Here we propose a new method to cool an object by illumination with a laser beam. Because of the directionality of laser beams and small spot sizes that can be achieved it is possible by the proposed method to create a heat sink that has an aperture for heat flow that matches the size of the device that needs to be cooled. The proposed method uses inelastic scattering as a means to remove heat from the material. If the anti-Stokes emission can be made more efficient than Stokes emission heat can be removed from the material. This can be achieved either by careful choice of the laser frequency with respect to the energy levels of the medium or by utilizing the using structures such as cavities and photonic crystals in which the density of states is such that favors emission in the anti-Stokes frequency than in the Stokes frequency. While the cooling effect is small it could be be sufficient to cool micro and nanoscale devices.

2. **Theory**

To explain the basic idea we'll consider a three level system (Figure 1). An optical beam with frequency $\omega_L$, which doesn't match the resonant frequencies of the medium ($\omega_{21}$ and $\omega_{32}$) cannot be absorbed. If the frequency $2\omega_L$ doesn't match the resonant frequencies in the medium simultaneous absorption of two photons cannot occur either. It can however be inelastically scattered when the selection rules do not forbid this process. Two processes can take place. First, if the system is in the ground state (1) it can make a transition to the excited state (2) by absorbing a photon of frequency $\omega_L$ and simultaneously emitting a photon of frequency $\omega_S = \omega_L - \omega_{21}$. This process is called Stokes scattering. Second, atoms in the excited state (2) can make a transition to the ground state by absorbing a photon of frequency $\omega_L$ and simultaneously emitting a photon at frequency $\omega_{AS} = \omega_L + \omega_{21}$. This process is called anti-Stokes scattering.

In a solid, if electromagnetic radiation is incident and the solid is transparent in that frequency then inelastic scattering will occur and the frequency $\hbar\omega_{21}$ could be the energy required to excite a quantum of atomic vibrations (also known as a phonon). In a gas $\hbar\omega_{21}$ could be the frequency necessary to excite an electron, molecular vibrations or molecular rotations.

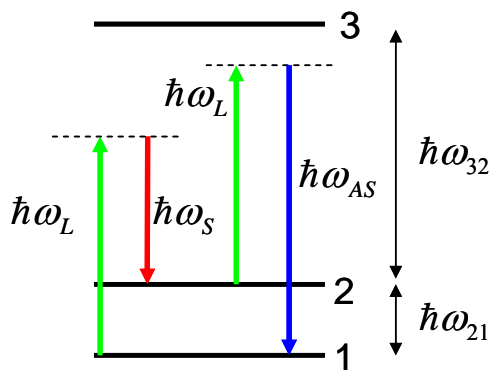

Figure 1 : Basic Raman scattering processes in a three- level system.

It is clear that in the Stokes process since the atom is excited in a higher energy level an amount of energy $\hbar\omega_{21}$ is deposited in the system (the so called quantum defect). In the anti-Stokes process since the system is de-excited an amount of energy $\hbar\omega_{21}$ is removed from the system. If the temperature of the system is $T$, then the ratio between the probability of finding the system in the excited state to the probability of finding the system on the ground state is approximately equal to $p_2/p_1 = e^{-\hbar\omega_{21}/kT}$. Since the system is most likely to be in the lower state, the rate of energy deposition (Stokes process) is much higher than the rate of energy removal (anti-Stokes process). Therefore the laser beam always results in the heating of the system.

We contend however, that the Stokes process can be suppressed or even eliminated and the anti-Stokes emission can be enhanced to the extent that heat is removed. We offer the following approaches to achieving this: (i) use of laser frequency below the vibrational frequency of the object, (ii) using the (quadratic) frequency dependence of the density of electromagnetic modes in a 3-D system, (iii) using a photonic crystal, (iv) using a electromagnetic cavity, (v) by matching the orientation of the electric polarization that is induced by the laser to the orientation of the cavity (vi) use of resonance enhancement of anti-Stokes scattering. These approaches will be detailed in the next section. It should be noted that for this approach to be successful, the rate of anti-Stokes emission must be much higher than the rate of linear and two-photon absorption processes; unwanted effects that will heat the object. Therefore it is preferred that the material is transparent at the pump frequency and also the pump frequency be such that the two-photon absorption cannot take place.

### 3. Implementation methods

### A. Elimination of Stokes emission by choice of pump frequency

If the laser frequency $\omega_L$ is lower than $\omega_{21}$ then the incoming photon energy is not enough to excite the $2 \rightarrow 1$ transition (refer to figure 2). Therefore the Stokes process is eliminated. In the case of device utilizing scattering in a solid $\hbar\omega_{21}$ is in the range 10 – 50 THz and if the scattering medium is a gas then in the case of electronic excitations $\hbar\omega_{21}$ can be in the optical part of the spectrum.

### B. Suppression of Stokes emission by choice of pump frequency

In this case the laser frequency $\omega_L$ is slightly higher than $\omega_{21}$. The Stokes process takes place, but the rate of the process can be much lower than the rate of anti-Stokes emission. The reason is that the spontaneous efficiency depends on the Stokes/anti-Stokes frequency. When $\omega_{21} << \omega_L$, the Stokes and anti-Stokes frequencies are very close and the difference in efficiency due to the frequency difference is negligible. Therefore the Stokes process is more efficient in that case. Now let's examine in detail what happens when $\omega_{21}$ is slightly smaller than $\omega_L$.

We first introduce two quantities of interest. The first, $S$, is the rate at which the spontaneous scattering process occurs per incident photon of frequency $\omega_L$, ($S_S$ for the Stokes and $S_{AS}$ for the anti-Stokes process). The second is the rate at which stimulated scattering process occurs, $g$ ($g_S$ for the Stokes and $g_{AS}$ for the anti-Stokes). The first quantity is proportional to the second. This is called the Einstein relationship and reads [2] : $S_S = g_S \times N_{BB}(\omega_S)$ and $S_{AS} = g_{AS} \times N_{BB}(\omega_{AS})$ where the proportionality constant equals

$$N_{BB}(\omega) = \frac{\omega^2}{\pi^2 c^3}\left(1 + \frac{1}{\exp(\hbar\omega/kT)-1}\right)$$

The factor $n_{TH}(\omega) = \frac{1}{\exp(\hbar\omega/kT)-1}$ is the occupation function of thermal photons (or thermal bosons in general) of frequency $\omega$. The constants $g_S$, $g_{AS}$ have a frequency dependence themselves, which is partly due to the probability that the energy level 2 is occupied. If level 2 is a vibrational level then $g_S = g'_S \omega_S (1 + n_{TH}(\omega_{21}))$ and $g_{AS} = g'_{AS} \omega_{AS} n_{TH}(\omega_{21})$. In these two expressions $n_{TH}(\omega_{21})$ is the occupation function for thermal phonons which has the same form as the occupation function for thermal photons that we gave above. The constants $g'_S$, $g'_{AS}$ are of the same order of magnitude as long as the laser frequency $\omega_L$ is not very close to one of the frequencies $\omega_{21}$ and $\omega_{31}$. In what follows we assume $g'_S = g'_{AS}$.

Let's consider the following example in order to see what kind of anti-Stokes enhancement one can expect. Consider a solid with a typical vibrational frequency $\omega_{21} \sim 2 \times 10^{13} Hz$ and decay rate for the vibrations $\gamma \sim 10^{11} Hz = 0.42 meV$. Because of this decay rate the optical absorption peak at the vibrational frequency has linewidth of $2\gamma$. To avoid absorption we therefore take $\omega_L - \omega_{21} = 10\gamma$. If the system is at room temperature then $1 + n_{TH}(\omega_{AS} = \omega_L + \omega_{21}) \cong 1.0014$ and $1 + n_{TH}(\omega_S = \omega_L - \omega_{21}) \cong 1 + \frac{kT}{\hbar(\omega_L - \omega_{21})} = 7.1$ which gives for the ratio of the efficiencies: $\frac{S_{AS}}{S_S} \cong \left(\frac{2\omega_L}{10\gamma}\right)^3 \frac{10\gamma}{kT} n_{TH}(\omega_{21}) = 400$. The net rate of transitions 2→1 that the system makes is $dn/dt = S_{AS} - S_S$ and the system will be losing energy until a temperature $T_f$ is reached at which $S_{AS} = S_S$. This will happen in our example when $(T/T_f)\exp(\hbar\omega_{21}(T^{-1} - T_f^{-1})) = 0.0025$ where $T_f$ is the final temperature, which is found to be ~ 105K.

There are certain points that the above discussions illustrate when choosing parameters for the cooling process for cooling with an incident frequency which is smaller or slightly greater than the Stokes shift. In both cases, the excitation energy $\hbar\omega_{21}$ must not be much higher that $kT$, where $T$ is the initial temperature of the material, because a high excitation energy means that the number of states of that energy to be occupied will be low. Therefore the amount of heat one can take out is limited. Therefore for a given temperature there is a given Stokes shift (excitation energy) which is optimum for the cooling process in the sense that it offers the maximum temperature lowering to be achieved. In the case of incident frequencies higher than the Stokes shift, it is better to select a value as close as possible to the Stokes shift but without suffering of linear absorption.

We can be more explicit about the frequency dependence of the coefficients $g'_S$, $g'_{AS}$ which we so far considered to be approximately equal. A quantum mechanical calculation for the probabilities per unit time for emission of Stokes and anti-Stokes radiation gives [2] :

$$g'_S \sim \frac{1}{\left((\omega_{31}-\omega_L)^2+\gamma_3^2\right)\left((\omega_{21}-\omega_L+\omega_S)^2+\gamma_2^2\right)}$$

$$g'_{AS} \sim \frac{1}{\left((\omega_{32}-\omega_L)^2+\gamma_3^2\right)\left((-\omega_{21}-\omega_L+\omega_{AS})^2+\gamma_2^2\right)}$$

In the above expressions it is assumed that the relaxation rates for the levels 3 and 2 are non-zero and equal to $\gamma_3$ and $\gamma_2$ respectively. Selecting $\omega_L-\omega_S=\omega_{21}$ and $\omega_{AS}-\omega_L=\omega_{21}$ we can compare the Stokes and anti-Stokes gain coefficients at the center of the Stokes and anti-Stokes Raman peaks :

$$\frac{g'_{AS}}{g'_S} \sim \frac{(\omega_{31}-\omega_L)^2+\gamma_3^2}{(\omega_{32}-\omega_L)^2+\gamma_3^2}$$

This shows that in the case where $\omega_L \approx \omega_{32}$ but $\omega_L < \omega_{31}$ the probability for anti-Stokes generation is enhanced by a factor :

$$\frac{g'_{AS}}{g'_S} = \frac{(\omega_{31}-\omega_L)^2+\gamma_3^2}{(\omega_{32}-\omega_L)^2+\gamma_3^2} \tag{1}$$

### C. Enhancement of anti-Stokes emission by means of resonant scattering

This approach to enhance the anti-Stokes efficiency relative to the Stokes is suggested by equation (1) of the previous section. Specifically one can have the pump frequency $\omega_L$ approach the frequency $\omega_{32}$, so that the frequency difference $\omega_{32}-\omega_L$ is low enough to make a difference in the anti-Stokes emission probability but high enough not to induce absorption of te pump beam. So if we set $\omega_{32}-\omega_L=10\gamma_3$ (1) gives :

$$\frac{g'_{AS}}{g'_S} = \frac{(\omega_{21}+10\gamma_3)^2+\gamma_3^2}{101\gamma_3^2}$$

In the case of a material with the property $\omega_{21} \gg 10\gamma_3$ the relative enhancement of the anti-Stokes gain coefficient is $g'_{AS}/g'_S \sim (\omega_{21}/10\gamma_3)^2$.

### D. Suppression of Stokes emission by means of a cavity and/or polarization control

The density of electromagnetic modes can be altered by means of a cavity. Specifically, it is possible to effect a higher mode density at the anti-Stokes frequency and lower at the stokes frequency. Since spontaneous anti-Stokes (Stokes) scattering occurs for every electromagnetic mode at the anti-Stokes (Stokes) frequency, the presence of a cavity can enhance the emission at the anti-Stokes frequency relative to the emission at the Stokes frequency.

As an example of cavities that modify the spontaneous emission rate of an oscillating dipole moment, we can look at a planar cavity (confinement only in 1 direction) formed by two perfectly reflecting planar mirrors (figure 2). In this case a dipole that is placed in the cavity equidistantly from the mirrors will have a spontaneous emission rate that depends on its orientation in the cavity. The emission rate modification factor is shown for two different dipole orientations, parallel and perpendicular to the cavity.

Another example shown in figure 3 is the cylindrical waveguide surrounded by a perfect metal surface. In the case of this waveguide the confinement of the optical waves is in 2 dimensions and the mode spectrum has stronger frequency dependence than the case of a planar cavity. The mode spectrum of a large volume is also shown for comparison.

The radiation emitted in the Stokes(anti-Stokes) frequency when a pump beam is incident on the material can be thought as produced by a polarization source oscillating at the Stokes (anti-Stokes) frequency and induced in the material by the pump beam. Therefore the mode spectra for dipoles centered in the center of the cavity shown in figures 2 and 3 are also relevant for the radiation produced in Raman scattering from a material volume placed in the center of the cavity (a thin material sheet in the 1$^{st}$ case or a long thin cylindrical tube in the 2$^{nd}$ case).

The selection rules for Raman scattering in materials essentially determine, for a given pump polarization incident in the material, the direction of the polarizations that radiate the Stokes and anti-Stokes waves.

In exploiting cavity mode spectra shown in Figures 2 and 3, the laser frequency is chosen such that the mode density at the anti-Stokes frequency is much higher than that at the Stokes frequency. This can result in a net absorption of phonons and hence the cooling of the object.

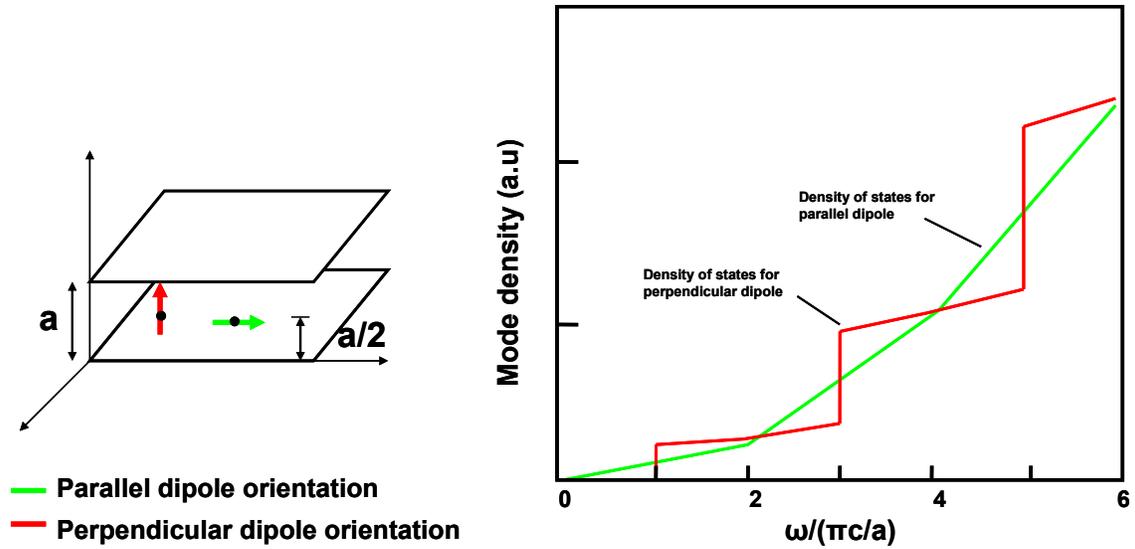

Figure 2 : Density of electromagnetic modes contributing to radiation from dipole placed midway in the planar cavity shown to the left(after [3]) . Both the density of states for a dipole oriented perpendicular (red) and parallel (green) to the cavity mirrors are shown separately.

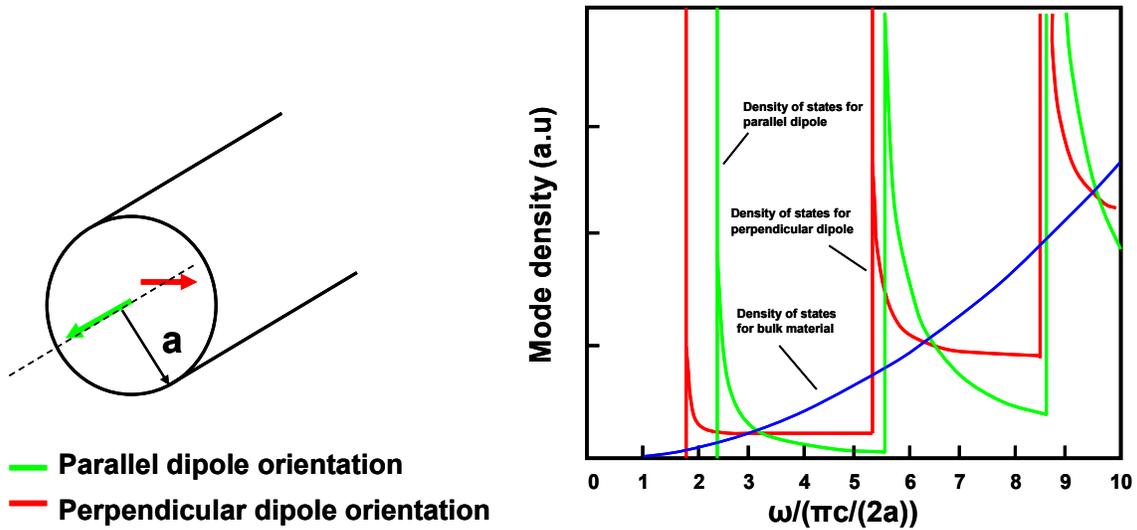

Figure 3 : Density of electromagnetic modes contributing to radiation from a dipole placed in the center of a the circular waveguide cavity shown on left. The density of states contributing to the radiation of the dipole placed in the centre of the waveguide at two different orientations is shown. The density of states of an infinite volume is also shown for comparison (after [3]).

### E.  Suppression of Stokes emission by means of a photonic crystal

In periodic structures with periodic dielectric characteristics, also known as photonic crystals, the Stokes process can be eliminated while at the same time the anti-Stokes process is enhanced. The reason is that the number of electromagnetic modes of propagation in a photonic crystal around a given optical frequency can be engineered. The number of modes can be increased, decreased or even made zero so that no propagation is allowed for waves of a given frequency.

The inhibition of a spontaneous emission process by such means was first proposed by Yablonovitch et al [4] as a mean to create low threshold lasers. In that work it was contemplated that spontaneous emission resulting from electronic transitions in a diode laser can be suppressed using a periodic structure. We herefore propose the use of a photonic crystal to eliminate the Stokes emission while at the same time enhancing the anti-Stokes emission. In figure 4 below an example of a photonic crystal structure studied by Yablonovitch et al [5]. The dispersion relationship for the waves that can propagate in the structure is also shown. The shaded area markes a set of frequencies where propagation along the given direction is impossible. This particular crystal exhibits the frequency bandgap also for all other directions of propagation.

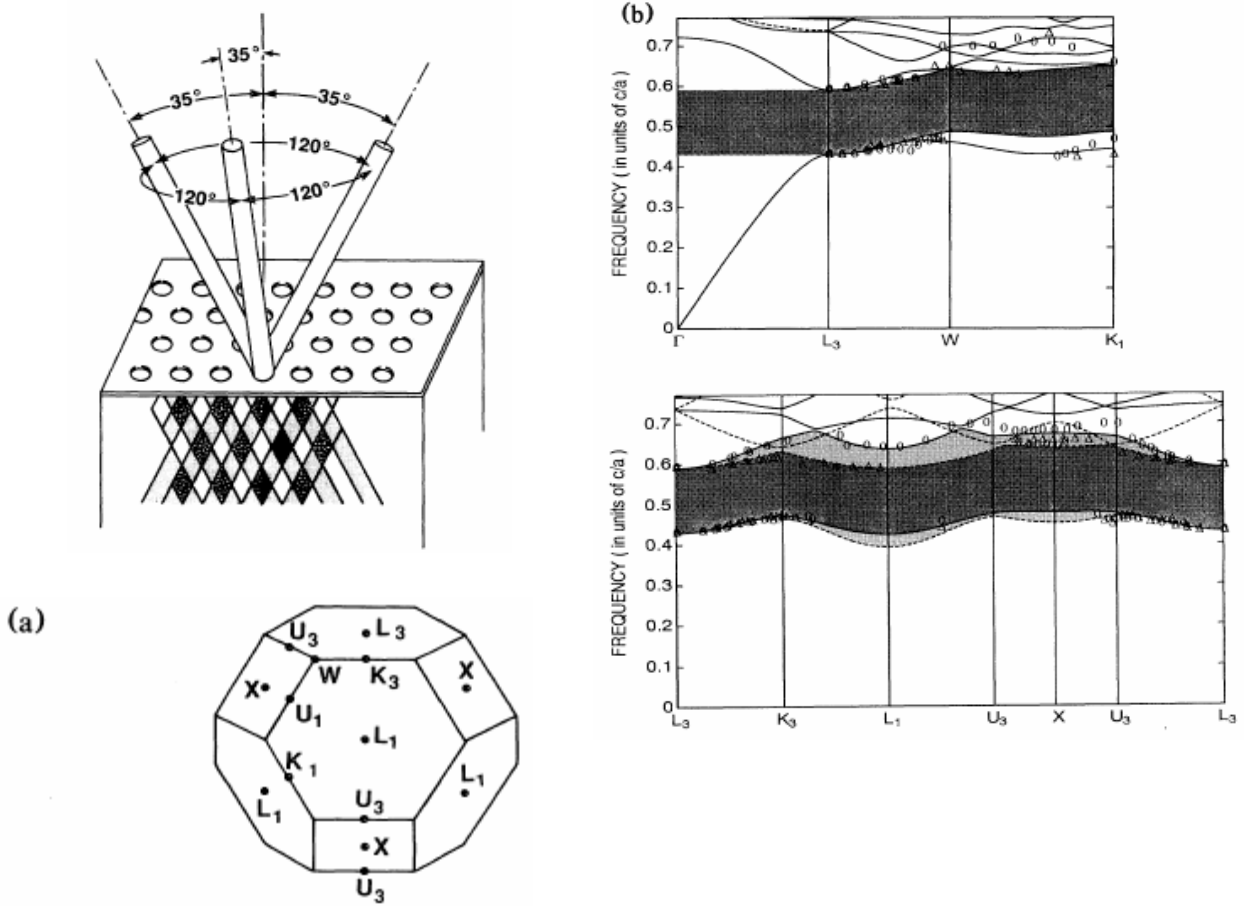

Figure 4 : An example of a photonic crystal is shown (after [5]). The photonic structure is formed by drilling equidistantly in the material series of holes in the directions shown in the schematic. In (a) the cell of periodicity is shown in reciprocal space and (b) shows for given the allowed modes of propagation along given direction in the Brillouin zone for different values of the photon energy. The shaded areas mark a region of frequencies where propagation with the given wavevectors is not possible. The particular crystal actually has a full bandgap. If other directions of propagation where shown for a range of frequencies propagation in any direction is impossible.

### 4. Summary

In summary, we have outlined several methods that can effect cooling on an object by illumination with a laser beam. All of the proposed methods aim to make anti-Stokes scattering more efficient that Stokes scattering. This can be achieved either by careful choice of the laser frequency with respect to the energy levels of the medium or by utilizing the using structures such as cavities and photonic crystals in which the density of states is such that favors emission in the anti-Stokes frequency than in the Stokes frequency. The cooling effect is small but could be sufficient to cool for example nano-scale devices.

**References**

[1] *http://www.intel.com/technology/silicon/power/index.htm*

[2] Quantum electronics, A. Yariv, John Wiley & sons, 3$^{rd}$ edition (1989)

[3] Optical processes in microcavities, Advanced series in applied physics, Eds, R.K. Chang, A.J. Campillo, World Scientific Pub. Co. (1996)

[4] "Inhibited spontaneous emission in solid-state physics and electronics", E. Yablonovitch, American Phys. Soc., Phys. Rev. Let. **58**, 2059 (1987)

[5] "Photonic band structure: the face-centered-cubic case", E. Yablonovitch, T.J. Gmitter, American Phys. Soc., Phys. Rev. Let. **63**, 1950 (1989)